\documentclass{elsart}
\usepackage{graphicx,amssymb}
\journal{PHYSICS LETTERS B}
\begin{document}
%
\begin{frontmatter}
\title{
Are protons still dominant at the knee of \\ the cosmic-ray energy spectrum?
}
\author[1]{M.Amenomori},
\author[2]{S.Ayabe},
\author[3]{D.Chen},
\author[4]{S.W.Cui},
\author[5]{Danzengluobu},
\author[4]{L.K.Ding},
\author[5]{X.H.Ding},
\author[6]{C.F.Feng},
\author[7]{Z.Y.Feng},
\author[8]{X.Y.Gao},
\author[8]{Q.X.Geng},
\author[5]{H.W.Guo},
\author[4]{H.H.He},
\author[6]{M.He},
\author[9]{K.Hibino},
\author[10]{N.Hotta},
\author[5]{Haibing Hu},
\author[4]{H.B.Hu},
\author[11]{J.Huang\corauthref{cor}},
\corauth[cor]{Corresponding author.}
\ead{huang@icrr.u-tokyo.ac.jp}
\author[7]{Q.Huang},
\author[7]{H.Y.Jia},
\author[12]{F.Kajino},
\author[13]{K.Kasahara},
\author[3]{Y.Katayose},
\author[14]{C.Kato},
\author[11]{K.Kawata},
\author[5]{Labaciren},
\author[15]{G.M.Le},
\author[6]{J.Y.Li},
\author[4]{H.Lu},
\author[4]{S.L.Lu},
\author[5]{X.R.Meng},
\author[2]{K.Mizutani},
\author[8]{J.Mu},
\author[14]{K.Munakata},
\author[16]{A.Nagai}
\author[1]{H.Nanjo},
\author[17]{M.Nishizawa},
\author[11]{M.Ohnishi},
\author[10]{I.Ohta},
\author[2]{H.Onuma},
\author[9]{T.Ouchi},
\author[11]{S.Ozawa},
\author[4]{J.R.Ren},
\author[18]{T.Saito},
\author[12]{M.Sakata},
\author[9]{T.Sasaki},
\author[3]{M.Shibata},
\author[11]{A.Shiomi},
\author[9]{T.Shirai},
\author[19]{H.Sugimoto},
\author[11]{M.Takita},
\author[4]{Y.H.Tan},
\author[9]{N.Tateyama},
\author[20]{S.Torii},
\author[21]{H.Tsuchiya},
\author[11]{S.Udo},
\author[4]{H.Wang},
\author[2]{X.Wang},
\author[6]{Y.G.Wang},
\author[4]{H.R.Wu},
\author[6]{L.Xue},
\author[12]{Y.Yamamoto},
\author[11]{C.T.Yan},
\author[8]{X.C.Yang},
\author[14]{S.Yasue},
\author[15]{Z.H.Ye},
\author[7]{G.C.Yu},
\author[5]{A.F.Yuan},
\author[9]{T.Yuda},
\author[4]{H.M.Zhang},
\author[4]{J.L.Zhang},
\author[6]{N.J.Zhang},
\author[6]{X.Y.Zhang},
\author[4]{Y.Zhang},
\author[4]{Yi.Zhang},
\author[5]{Zhaxisangzhu},
\author[7]{X.X.Zhou}, \\
(The Tibet AS$\gamma$ Collaboration)
\address[1]{Department of Physics, Hirosaki University, Hirosaki 036-8561, Japan}
\address[2]{Department of Physics, Saitama University, Saitama 338-8570, Japan}
\address[3]{Faculty of Engineering, Yokohama National University, Yokohama 240-8501, Japan}
\address[4]{Key Laboratory of Particle Astrophysics, Institute of High Energy Physics, Chinese Academy of
 Sciences, Beijing 100049, China}
\address[5]{Department of Mathematics and Physics, Tibet University, Lhasa 850000, China}
\address[6]{Department of Physics, Shandong University, Jinan 250100, China}
\address[7]{Institute of Modern Physics, South West Jiaotong University, Chengdu 610031, China}
\address[8]{Department of Physics, Yunnan University, Kunming 650091, China }
\address[9]{Faculty of Engineering, Kanagawa University, Yokohama 221-8686, Japan}
\address[10]{Faculty of Education, Utsunomiya University, Utsunomiya 321-8505, Japan}
\address[11]{Institute for Cosmic Ray Research, University of Tokyo, Kashiwa 277-8582, Japan }
\address[12]{Department of Physics, Konan University, Kobe 658-8501, Japan}
\address[13]{Faculty of Systems Engineering, Shibaura Institute of Technology, Saitama 337-8570, Japan}
\address[14]{Department of Physics, Shinshu University, Matsumoto 390-8621, Japan}
\address[15]{Center of Space Science and Application Research, Chinese Academy of Sciences, Beijing 100080, China }
\address[16]{Advanced Media Network Center, Utsunomiya University, Utsunomiya 321-8585, Japan}
\address[17]{National Institute for Informatics, Tokyo 101-8430, Japan}
\address[18]{Tokyo Metropolitan College of Aeronautical Engineering, Tokyo 116-0003, Japan}
\address[19]{Shonan Institute of Technology, Fujisawa 251-8511, Japan}
\address[20]{Advanced Research Institute for Science and
Engineering, Waseda University, Tokyo 169-8555, Japan}
\address[21]{RIKEN, Wako 351-0198, Japan}
\begin{abstract}
 A hybrid experiment consisting of emulsion chambers, burst detectors  and
the Tibet~II air-shower array was carried out at Yangbajing (4,300 m a.s.l.,
606 g/cm$^2$) in Tibet to obtain the energy spectra of primary protons 
and heliums. 
From three-year operation, these energy spectra are deduced
between $10^{15}$ and $10^{16}$ eV by
triggering the air showers associated with a high energy core
and using a neural network method in the primary mass separation.
The proton spectrum can be expressed by a single power-law function with
a differential index of $-3.01 \pm 0.11$ and $-3.05 \pm 0.12$ based on the 
QGSJET+HD and SIBYLL+HD models,
respectively, which are steeper than that extrapolated from the direct 
observations of $-2.74 \pm 0.01$ in the energy range below $10^{14}$ eV.
The absolute fluxes of protons and heliums are derived 
within 30\% systematic errors
depending on the hadronic interaction models
used in Monte Carlo simulation.
The result of our experiment suggests
that the main component responsible for 
the change of the power index of the all-particle spectrum around
$3 \times 10^{15}$ eV, so-called ``knee'', is composed 
of nuclei heavier than helium.
This is the first measurement of the differential energy spectra of 
primary protons and heliums by selecting them event by event at
the knee energy region.
\end{abstract}
\end{frontmatter}
\section{Introduction}

The energy spectrum of cosmic rays is described by a power-law function 
in a wide energy range from about $10^{10}$ eV to $10^{20}$ eV, 
however, it shows slight changes of the power-law index
at several points. These break points of the power-law spectrum 
are assumed to be related to the origin, acceleration mechanism
and propagation mechanism of cosmic rays in the Galaxy. 
One of the break points, in which the present paper
is concerned, is traditionally referred to as 
the ``knee'' located around 3 $\times$ $10^{15}$ eV. 
Many experiments have reported \cite{All-spectrum} that the power-law indices
below and above the knee approximately take the values $-2.7$ and 
$-3.1$, respectively. Although existence of the knee has been well established 
experimentally, there are still controversial arguments on its origin.
One of them is a possibility that acceleration  mechanism could be
less effective above the knee. Along with this line, there is a general
consensus that stochastic shock acceleration at supernova blast waves 
could explain the cosmic-ray spectrum up to about $Z \times 10^{14}$ eV \cite{Lagage},
or perhaps even higher to $Z \times 10^{15}$ eV \cite{Jokipii}, where $Z$ denotes the 
atomic number, despite lack of direct evidence. Another argument 
attributes the knee structure to 
the leakage of the cosmic rays from the galaxy \cite{Peters}.
It is noted that both scenarios mentioned above give a 
rigidity-dependent cutoff for each chemical component 
leading to a heavy-enriched composition of primary cosmic rays at the knee.
On the other hand, there is another approach
in which cosmic rays around and beyond the knee are assumed
to be of extra-galactic origin such as the active galactic nuclei 
\cite{Protheroe} or gamma-ray bursts \cite{Plaga}. In this case, the 
primary chemical composition is expected to become proton-enriched.
There have been some calculations of the primary cosmic-ray energy spectrum 
based on various models on the origin of the knee \cite{Horandel1},
but all of them are still under debate due to lack of detailed knowledge 
about the chemical composition around the knee.

Among primary cosmic rays, protons are the key component for understanding 
the origin of the knee. Direct measurements of primary cosmic rays on board 
balloons or satellites are the best ways, however, the energy region covered 
by them are limited up to $10^{14}$ eV. The chemical composition of primary 
cosmic rays around the knee, therefore, has been studied with ground-based 
air-shower experiments and/or air Cherenkov telescopes. Since the sensitivity 
to the mass separation among cosmic-ray nuclei with ground-based  experiments 
is limited, only gross features such as average mass number have been discussed. 
A lot of reports have so far been made on the energy spectrum as well as the 
chemical composition of primary cosmic rays, however, there are still serious 
disagreements among them especially on the chemical composition \cite{Horandel2}.

It is possible, however, to improve the sensitivity of an air shower experiment
to the primary cosmic-ray mass separation by adding a function to observe
the energy-flow characteristics of  air-shower cores at a high-mountain altitude.
The Tibet hybrid experiment \cite{Amenomori1} was designed to detect 
$\gamma$-families in an air shower core by large-area emulsion chambers
in coincidence with an accompanied air shower, where a bundle of energetic
$\gamma$-rays and electrons detected by the emulsion chambers are called a
$\gamma$-family, which is caused by a young air shower. 
Among primary cosmic rays, protons and heliums can penetrate deep into the atmosphere 
and produce a young air shower accompanied by a $\gamma$-family event most efficiently
due to their longest interaction mean free paths in the air. Therefore, tagging an air 
shower with a $\gamma$-family event enriches the proton and helium component naturally.
Another merit in doing a hybrid experiment in Tibet is that 
the atmospheric depth of the experimental site 
(4300 m a.s.l., 606 g/{cm}$^{2}$) is close to the maximum development 
of the air showers with energies around the knee almost irrespective of the masses 
of primary cosmic rays. We can determine the primary cosmic-ray energy 
much less dependently upon the chemical composition \cite{Amenomori2} 
than those experiments at the sea level. 
Thus, the Tibet hybrid experiment enables us to measure the primary 
proton and helium differential energy spectra by selecting them
event by event.
 
In this paper, we briefly report on the study of the energy spectra 
of the proton and helium components in cosmic rays around the knee
energy region obtained with the Tibet hybrid experiment.

\section{Experiment}

The Tibet hybrid experiment, consisting of emulsion chambers (ECs), burst 
detectors (BDs) and the Tibet-II air-shower array (AS), composed of 221 
scintillation counters each placed on a 15 m square grid with an enclosed area 
of 36,900 m$^{2}$ was operated at Yangbajing in Tibet during the period from 
November 1996 through August 1999 \cite{Amenomori1} and a total live time of
699.2 days. 

The AS is used to measure the shower size and the arrival direction of each air shower. 
Any four-fold coincidence in the detectors is used as the trigger condition 
for air-shower events. The air shower direction 
can be estimated with an error smaller than $1^\circ$. The primary energy of 
each event is determined by the shower size ($N_{\rm{e}}$). The energy resolution is 
estimated to be 17\% at energies around $10^{15}$ eV by our simulation, 
almost independent of the interaction models used.

The ECs and the BDs are constructed near the center of the AS \cite{Amenomori3}, 
and are used to detect high-energy air shower cores accompanied by air showers 
induced by primary cosmic rays with energies above $\sim 10^{14}$ eV. The total area
of ECs is 80 m$^{2}$. The basic structure of each EC is a multilayered
sandwich of lead plates and X-ray films (FUJI X-RAY FILM TYPE 200) of 40 cm $\times$ 50 cm in area, 
where X-ray films are put every 1.0 cm of lead in the chamber.
The X-ray films in ECs are replaced by new ones every year to reduce the background.
The ECs are used to measure the energy, the position and the arrival direction 
of each $\gamma$-family shower with energies above 1 TeV. 
The BD with the same area are placed just below 4 ECs, 
namely, 4 ECs are set above one unit of the BD. Thus, 400 blocks of ECs 
and 100 BDs in total are used in this experiment. 
A burst event is triggered when any two-fold coincidence of signals from four photodiodes 
of each BD appears. When the BDs trigger an event, its accompanying air shower is 
simultaneously recorded. The BDs are used to measure the burst size $N_{\rm{b}}$ and 
the position of each air shower core. The arrival direction of the $\gamma$-family event 
is determined by the spatial reconstruction of the cascade showers in ECs, whose details 
are described in \cite{Ozawa}. The matching between an AS and a BD event is made by their 
arrival time stamps, and the matching between a $\gamma$-family event in EC and 
the BD event is made by their positional correlation, and 
the matching between the $\gamma$-family event in EC and 
the AS event is made by their directional correlation. 

In the following analysis, we present our results based on the  
ECs and AS data, as those obtained from the BDs data were published in the 
previous paper \cite{Amenomori3}.

\section{Simulation}

We have carried out a detailed Monte Carlo (MC) simulation of air showers and 
$\gamma$-families using the simulation code  CORSIKA (version 6.030) 
including QGSJET01 and SIBYLL2.1 hadronic interaction models \cite{Heck1}.
From a point of view to check the dependence of the obtained results on 
the assumed primary cosmic-ray composition in MC, two primary composition 
models are examined as the input energy spectra, namely a heavy 
dominant (HD) and a proton dominant (PD) ones \cite{Amenomori1}.
The energy spectrum of each component in the HD model has
a rigidity-dependent break point of the power index with proton's
knee around $1.5 \times 10^{14}$ eV leading to the dominance of the heavy component 
at the knee energy region. On the other hand, it is assumed in the 
PD model that light components are dominant up to the knee, in which
every component has the same break point of the power index at 
the knee energy. In both models, the fraction and the power index of 
each component are determined by fitting to the fluxes of the elements 
obtained by direct observations below $10^{14}$ eV,
and fitting the sum of the each element at higher energies to the all
particle flux obtained by air shower experiments.
Therefore, the difference between two models exists in the fraction of the
elements above 10$^{14}$ eV.
The fractional contents of the assumed primary cosmic-ray flux models
are listed in Table~\ref{tab:01}, together with those 
for making air showers accompanied by $\gamma$-families, where
M denotes the sum of medium heavy elements between helium and iron.
One can see from Table~\ref{tab:01} that 100\% of the $\gamma$-family 
events are induced by protons and heliums below 10$^{15}$ eV.
At the higher energy range, however, the contribution of the other 
nuclei heavier than helium increases with the energy which amounts to 
around 15\% and 4\% in case of HD model and PD model, respectively. 
The method of the separation of primary mass groups using a neural
network method is described in the next section.

%
%
\begin{table}
\begin{center}
\caption{Fractions of the proton(P), helium(He), medium(M) and iron(Fe) components in the assumed primary 
cosmic-ray spectrum of the HD and PD models(upper table), together with those
for making air showers accompanied by $\gamma$-families (lower table)  
(see the Section 4).}
\begin{tabular}{lccccccccc}\hline
Primary   &        &   \multicolumn{4}{c}{HD}  &  \multicolumn{4}{c}{PD}     \\  
          &  Energy ( eV ) & P  & He & M & Fe & P & He & M & Fe    \\  
\hline
Generated &  $10^{14}$ - $10^{15}$ & 22.6 & 19.2 & 36.0 & 22.2 & 39.0 & 20.4 & 31.2 & 9.4 \\
(\%) & $10^{15}$  - $10^{16}$ & 11.0 & 11.4 & 38.5 & 39.1 & 38.1 & 19.4 & 32.6 & 9.9 \\ 
\hline
     &                       &     &    &    &  & & & & \\
\end{tabular}
\begin{tabular}{lcccccc}\hline 
Model &Energy(eV) &  P & He & M & Fe \\
\hline
QGSJET+HD& $10^{14}$-$10^{15}$ & 87.3$\pm$1.2 & 12.7$\pm$1.2 & 0 & 0 \\
(\%) & $10^{15}$-$10^{16}$ & 58.9$\pm$0.9 & 27.2$\pm$0.8 & 12.3$\pm$0.9 & 1.6$\pm$0.3 \\
\hline 
SIBYLL+HD& $10^{14}$-$10^{15}$ & 87.2$\pm$0.8 & 12.8$\pm$0.8 & 0 & 0 \\
(\%)& $10^{15}$-$10^{16}$ &  57.3$\pm$0.7 & 24.2$\pm$0.7 & 16.9$\pm$0.8 & 1.6$\pm$0.3 \\
\hline 
QGSJET+PD& $10^{14}$-$10^{15}$ & 91.8$\pm$0.8 & 8.2$\pm$0.8 & 0 & 0\\ 
(\%)& $10^{15}$-$10^{16}$ & 80.0$\pm$0.6 & 16.0$\pm$0.6 & 3.4$\pm$0.4 & 0.6$\pm$0.1\\
\hline 
SIBYLL+PD& $10^{14}$-$10^{15}$ & 94.2$\pm$0.6 & 5.8$\pm$0.6 & 0 & 0\\ 
(\%)& $10^{15}$-$10^{16}$ & 78.7$\pm$0.6 & 17.9$\pm$0.6 & 3.4$\pm$0.4  & 0.06$\pm$0.01 \\
\hline 
\end{tabular}
\label{tab:01}
\end{center}
\end{table}

The numbers of generated $\gamma$-family events satisfying the criteria
described in the next section
are 5252, 5926, 8588 and 7376 for the QGSJET+HD, QGSJET+PD, SIBYLL+HD and
SIBYLL+PD models, respectively, which are more than 30 times greater than
the experimental statistics. 


\section{Analysis}

Shower spots registered by high-energy $\gamma$-rays or electrons in the 
X-ray films were automatically analyzed  by using image scanners \cite{Ozawa}. 
The $\gamma$-family events are selected by imposing the
conditions of ${E_{\gamma}}^{th} = 4$ TeV, $N_{\gamma}\geq 4$, $\displaystyle{\sum_{i}{E_{\gamma}}^{i}}\geq
20$ TeV and $<R> \geq 0.2$ cm, where ${E_{\gamma}}^{th}$ is the minimum energy 
for a cascade shower, $N_{\gamma}$ the number of cascade showers in a $\gamma$-family, 
$\sum{E_{\gamma}}$ the sum energy of cascade showers in a $\gamma$-family and 
$<R>$ ($<R> = \displaystyle{\sum_{i}{r_{\gamma}}^{i}/N_{\gamma}}$) the mean lateral spread in a $\gamma$-family. 
In this experiment, we observed a total of 177 $\gamma$-family events,
each of which is accompanied by an air shower with $N_{\rm{e}} > 2 \times 10^5$ corresponding 
approximately to $ 5 \times 10^{14}$ eV for a proton.

The separation of the primary mass is made with use of
a feed-forward  artificial neural network (ANN \cite{Lonnblad})
whose applicability to our experiment was well confirmed by the 
Monte Carlo simulation \cite{Zhang1,Amenomori3}.
The first task of the ANN is to separate protons from everything 
else by training the network with a proton flag.
As shown in Table~\ref{tab:01}, heliums are the main component
of the contamination at lower energies while
medium heavy component also contributes in higher energies
depending on the primary composition.
Then the second task of ANN is to separate a group of proton+helium 
from others above 10$^{15}$ eV by training the network with 
a light element flag, i.e., a flag for proton or helium.
Thus we can obtain helium spectrum by subtracting the number of 
proton events obtained in the first task from the proton+helium dataset.

We examine four cases, QGSJET + HD, QGSJET + PD, SIBYLL + HD and SIBYLL + PD models.
The following six parameters are input to the ANN  with 30 hidden 
nodes and 1 output unit, which are abbreviated to a $6:30:1$ network
\footnote{We compared the results from various number of hidden nodes
and found the results are not changed when we use the number of hidden
nodes greater than 20.}
:
(1) $\sum E_{\gamma}$ in ECs, 
(2) $N_{\gamma}$ in ECs, 
(3) $<R>$ in ECs,
(4) the mean  energy-flow spread  $<ER> = \displaystyle{\sum_{i}{E_{\gamma}}^{i}{r_{\gamma}}^{i}/N_{\gamma}}$ 
in ECs, where ${E_{\gamma}}^{i}$ and ${r_{\gamma}}^{i}$ are the energy of each cascade shower in 
the $\gamma$-family and 
its distance from the energy-weighted family-center, respectively,
(5) the air shower size  $N_{\rm{e}}$ in AS,
and (6) $\sec\theta$, where $\theta$ is the zenith angle of the arrival
direction.

\begin{figure}[t]
\begin{center}
\includegraphics*[width=7.5cm]{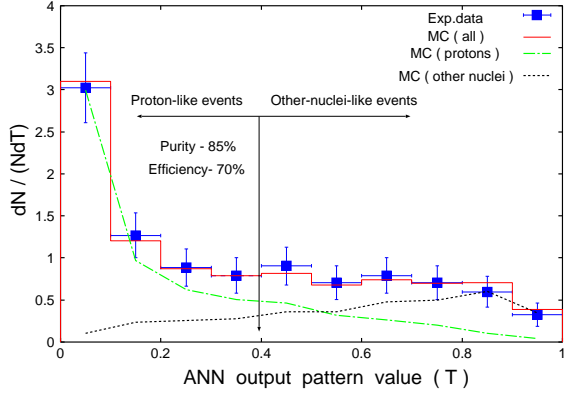}
\end{center}
\caption{ANN output pattern value ($T$) distributions 
compared with MC (QGSJET+HD model).
Used symbols are ; experimental data (the closed squares), 
MC:all (the solid-red histogram),
MC:protons (the dash-dotted green line), 
MC:other nuclei (the dotted-black line). 
}
\label{fig:1}
\end{figure}

One MC data set was generated for each of the four models and 
divided into two subsets; the one is used for training the ANN, 
and the other for estimating the ability of the ANN to classify 
the nuclear species. Then, the training MC data subset is fed 
to the ANN in a number of training cycles of 2000. 
\footnote{The training curve 
becomes stable at number of training cycles 500.} \ To train 
the ANN in separating protons from others, the input patterns 
for protons and other nuclei are set to 0 and 1, respectively. 
After the training, the another MC data subset is used to estimate 
the purity and the selection efficiency of protons.
Then, the ANN output pattern value ($T$) is a real number from 0 to 1.
The $T$ distributions in the QGSJET+HD model is presented 
in Fig.~\ref{fig:1}, together with the experimental data. 
One can see that the experimental data is in a good agreement with the 
MC prediction, and that the proton-induced
events are clearly separated from other nuclei. 
\footnote{We also examined the QGSJET+PD, SIBYLL+HD and SIBYLL+PD models, 
obtaining similar results to the ones as shown in Fig.~\ref{fig:1}, for 
the selected events are already mostly proton and helium origin
in all models as shown in Table~\ref{tab:01}.
Of course there are some differences between HD model and PD model,
reflecting the degree of the contamination, however, it is not large enough
to rule out one of them because of experimental statistical errors.
This figure demonstrates how the proton-induced events are 
separated from those induced by other nuclei.}
%
%
We define a critical value of $T$ to separate protons from others
requiring the high purity, which reduces the effect of the
contamination, and the high selection efficiency of the proton
events, which reduces the statistical error.
Since these two factors are of the competing nature,
the purity and the selection efficiency are calculated
as a function of the critical value $T_{c}$ and its value is set
as 0.4 where average purity and selection efficiency 
over whole energy range are greater than
approximately 85\% and 70\% for all models, respectively. 
It was confirmed that the final result does not depend
on the different choice of the $T_{c}$ around 0.4.
The high value of the purity 
as listed in Table~\ref{tab:02} and their mutual deviation among different models
in the same energy interval being within 4\% assures us the quality of the 
proton selection.
%
%
%
\begin{table}
\begin{center}
\caption{The purity of the selected events by $T<0.4$.}
\begin{tabular}{lcccc}\hline 
Model &Energy(eV)&\multicolumn{2}{c}{Purity(\%)} \\
& &  HD&PD \\
\hline
QGSJET& $10^{14}$-$10^{15}$ & 96.7$\pm$0.7 & 97.4$\pm$0.4  \\
 & $10^{15}$-$10^{16}$ & 83.1$\pm$1.6 &86.7$\pm$0.8\\
\hline 
SIBYLL& $10^{14}$-$10^{15}$ & 96.2$\pm$0.5 &97.3$\pm$0.3  \\
& $10^{15}$-$10^{16}$ &  82.8$\pm$1.2 &86.1$\pm$0.7 \\
\hline 
\end{tabular}
\label{tab:02}
\end{center}
\end{table}

Finally, we obtained 111, 111, 112, 112 proton-like events  out of 177 
observed events after the $\gamma$-family selection  based on the  
QGSJET+HD, QGSJET+PD, SIBYLL+HD and SIBYLL+PD models, respectively,
among which 110 events are identical and one event belongs 
only to QGSJET analysis and two events belong only to SIBYLL analysis. 

\section{Results and Discussions}

\begin{figure}[t]
\begin{center}
\begin{minipage}[b]{8.0cm}
\includegraphics*[width=7.5cm]{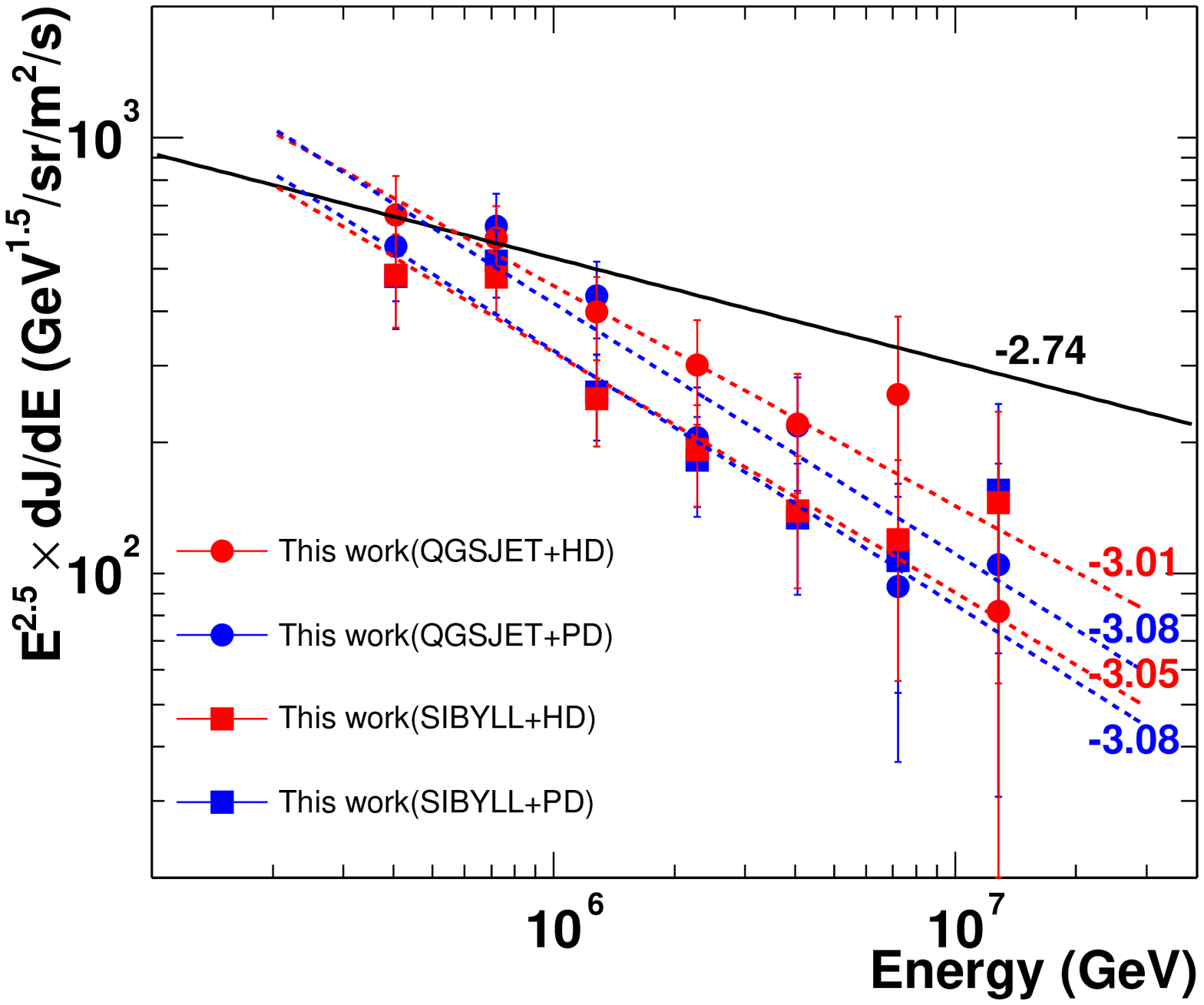}
\end{minipage}%
\begin{minipage}[b]{8.0cm}
\includegraphics*[width=7.5cm]{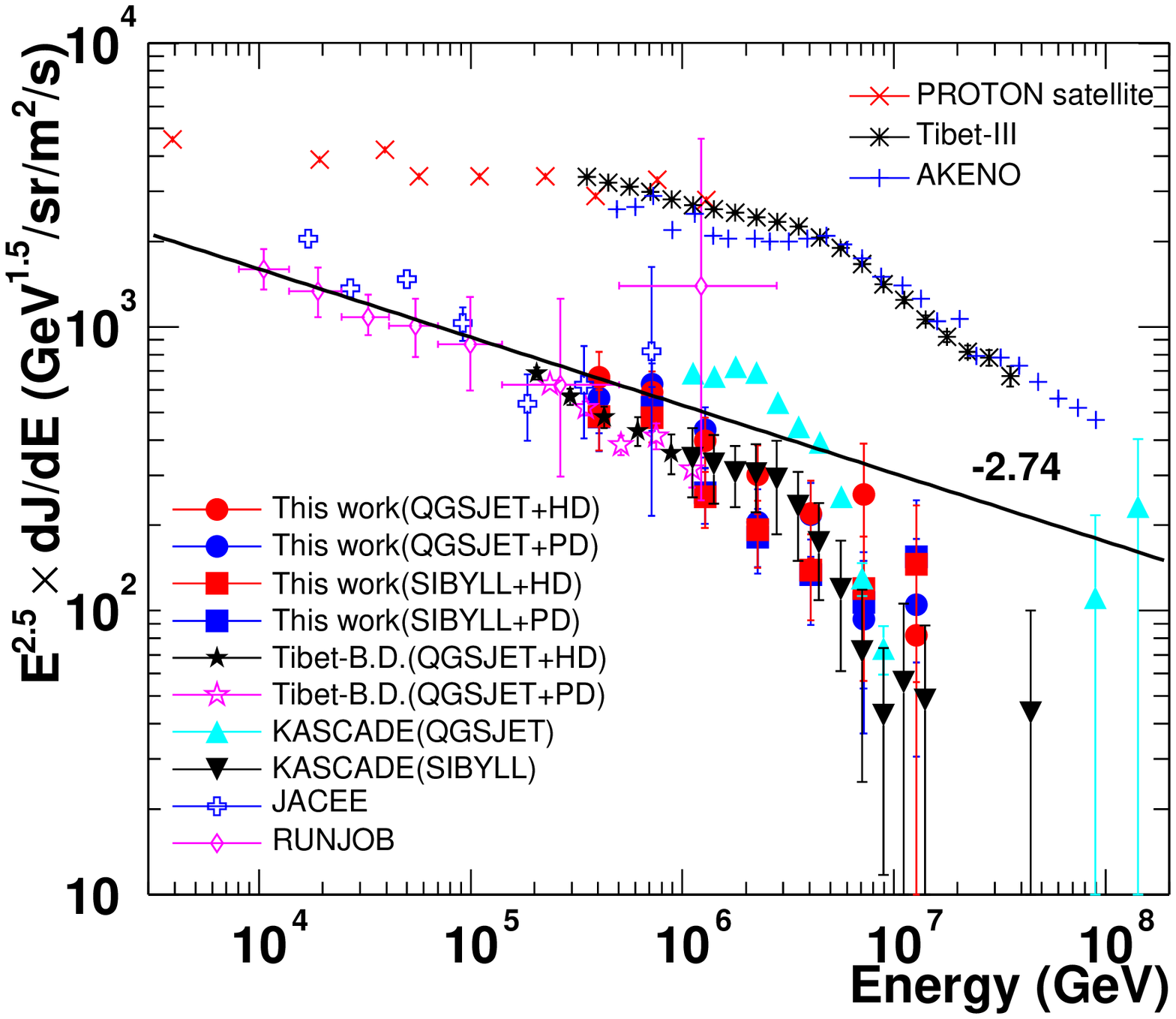}
\end{minipage}
\end{center}
\caption{Energy spectra of primary cosmic-ray protons obtained by the present 
experiment (a) and they are compared with other experiments (b):
Tibet-B.D.\cite{Amenomori1}, KASCADE\cite{Kampert}, JACEE\cite{Asakimori1} and 
RUNJOB\cite{Apanasenko1}. The all-particle spectra are from the experiments : 
PROTON satellite\cite{Grigorov}, Tibet-III\cite{Amenomori4} and 
AKENO\cite{Nagano1}. 
For the solid line with the power index $-2.74$,
see the text.}
\label{fig:2}
\end{figure}

In Fig.~\ref{fig:2}, we show the measured primary cosmic-ray proton energy 
spectra assuming the two interaction models (QGSJET and SIBYLL) and
two primary composition models (HD and PD), together with the results 
from other experiments.  As seen in Fig.~\ref{fig:2}, the present 
results assuming the HD and PD  models in the simulation 
are in a good agreement with each other within the statistical errors. 
The measured proton energy spectra can be expressed by a single 
power-law function of a differential form
$J(E)({\rm{m}}^{-2}{\cdot}{{\rm{s}}^{-1}}{\cdot}{{\rm{sr}}^{-1}}{\cdot}{{\rm{GeV}}^{-1}}) 
= A \times {10^{-13}} \times (\frac{E}{{10}^{6} GeV})^{-B}$,
where ($A$,$B$) is ($4.56 \pm 0.46$, $3.01 \pm 0.11$), ($4.14 \pm 0.44$, $3.08 \pm 0.11$),
($3.21 \pm 0.34$, $3.05 \pm 0.12$) and ($3.24 \pm 0.34$, $3.08 \pm 0.12$) 
based on the  QGSJET+HD, QGSJET+PD, SIBYLL+HD and SIBYLL+PD models, 
respectively, where the errors quoted are the statistical ones. 
The error in the spectral index is statistics dominant, 
while that in the absolute flux value is model-dependence dominant. 
For the absolute flux value, the QGSJET model gives approximately 
30\% higher flux than the SIBYLL model. 
This can be mainly attributed to the difference of Feynman 
$x_{F}$-distribution of charged mesons between QGSJET and SIBYLL model 
in the very forward region at a collision \cite{Heck1}. The Feynman 
$x_{F}$-distribution in the SIBYLL model is harder than that 
in the QGSJET model in the $x_{F} > 0.2$ region, so that the 
generation efficiency of $\gamma$-families by the former model 
becomes higher than the latter, resulting in a lower proton flux 
in the case of the SIBYLL model.
As compared in Fig.~\ref{fig:2}, the present results are consistent with 
those obtained by the burst detectors in this experiment within 25\% 
\cite{Amenomori3}. 
This implies that the systematic energy-scale uncertainty 
in our experiment is estimated to be 10\% level.
A solid straight line with the power index $-2.74$ drawn in 
Fig.~\ref{fig:2} is the best fitted line for the data points 
in the energy region below $10^{14}$ eV observed by 
recent direct measurements \cite{Gaisser}, which is harder than the indices
of our proton spectra.

%
%
%

%
%
\begin{figure}[t]
\begin{center}
\begin{minipage}[b]{8.0cm}
\includegraphics*[width=7.5cm]{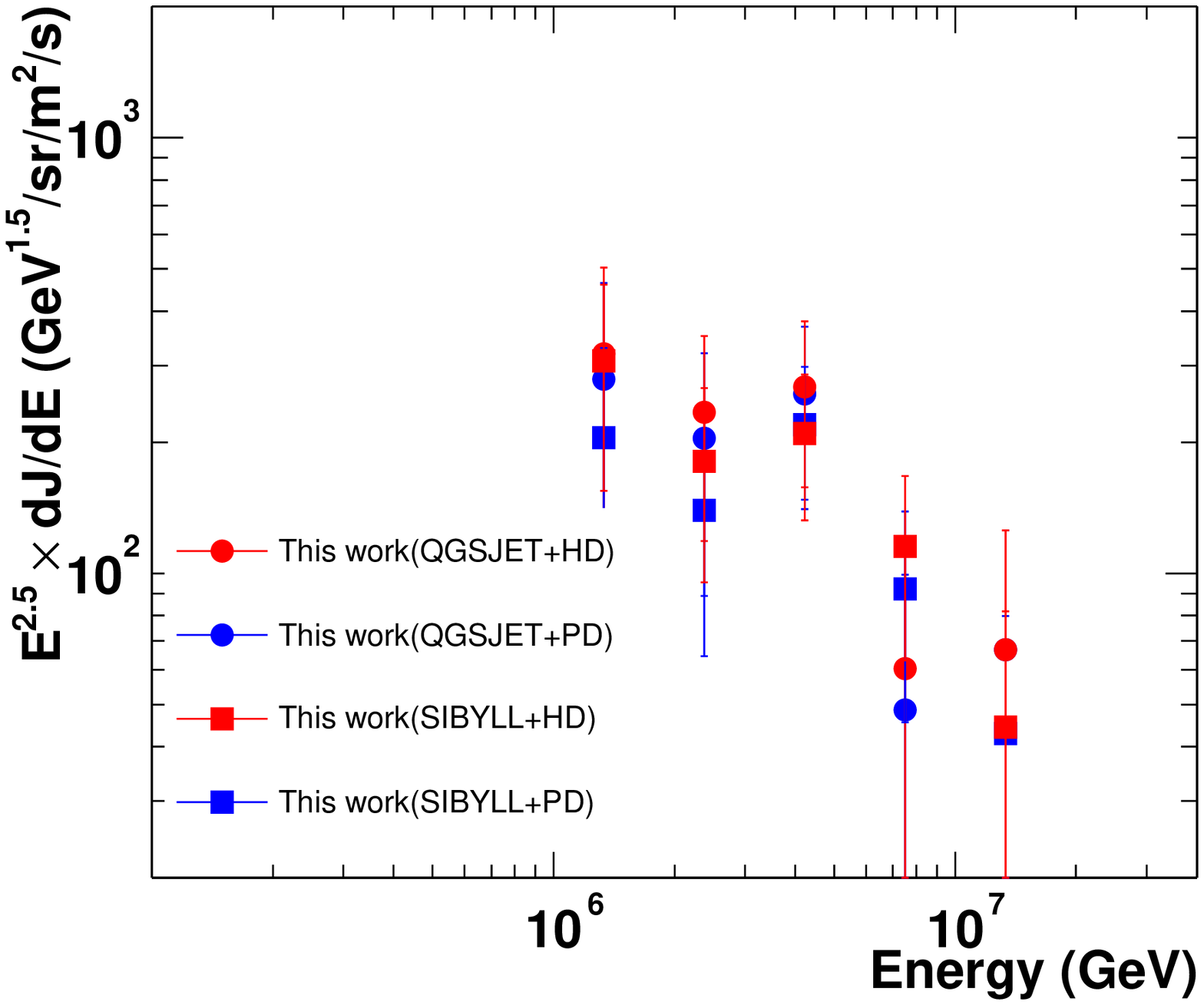}
\end{minipage}%
\begin{minipage}[b]{8.0cm}
\includegraphics*[width=7.5cm]{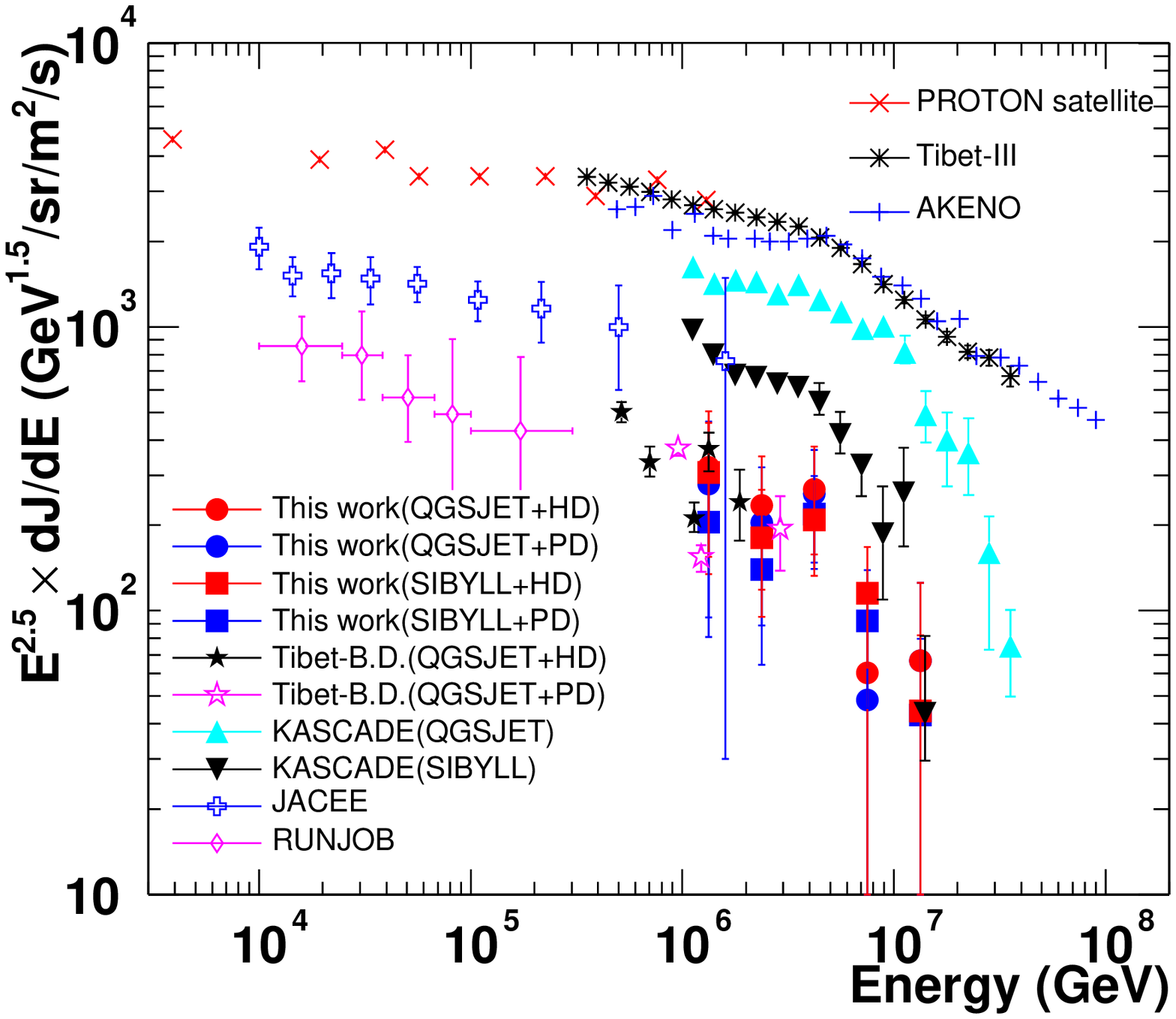}
\end{minipage}
\end{center}
\caption{Energy spectra of primary cosmic-ray helium nuclei obtained by 
the present experiment (a) and they are compared with other experiments (b)}
\label{fig:3}
\end{figure}

Thanks to its light mass, the helium component can also
trigger our hybrid experiment although the efficiency at $10^{15}$ eV
is about 4 times lower than the case of protons. 
The ANN method is again applied to obtain the helium spectrum over the 
energy $10^{15}$ eV.
Because of the training algorithm of ANN, it is not possible to train
the network to separate heliums from others directly, for the helium mass is
between protons and other heavy nuclei and 
the characteristics of the helium event is smeared out by
the fluctuation tail from the both sides.
Therefore we train the network to separate light component 
(proton or helium) from other nuclei, by assigning 0 to light component
and 1 to other nuclei.
The critical value $T_{c}$ to select light component is set as
0.2 where the selection efficiency reaches to 70\% and the
purity is 93\% for all models.
Then, the helium spectra can be obtained by subtracting the number of
protons, which are previously obtained by proton-training,
from the number of proton+helium events.
Above mentioned procedure was applied on each energy bin to obtain the
energy spectra of heliums and the result is shown 
in Fig.~\ref{fig:3}, where the same dependence of the absolute
intensity on the interaction models is seen as in the case of proton spectra.

%
%
\begin{figure}[t]
\begin{center}
\includegraphics*[width=7.5cm]{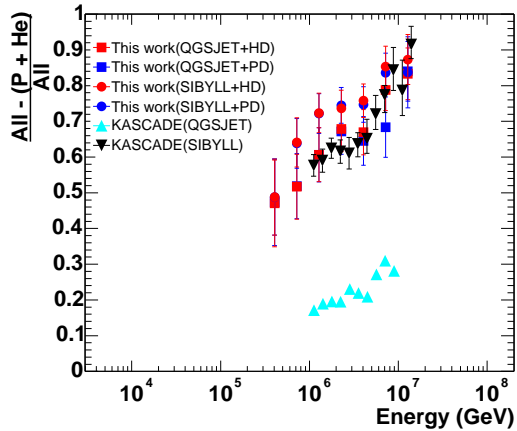}
\end{center}
\caption{Fraction of the primary cosmic-rays  heavier than helium nuclei 
obtained by assuming the QGSJET and SIBYLL interaction models. Our results 
are compared with those by the KASCADE experiment\cite{Kampert}.}
\label{fig:4}
\end{figure}

We can also estimate the fraction of the nuclei heavier than helium 
in cosmic rays around the knee using the proton+helium spectra and the 
all-particle energy spectrum obtained by the Tibet air shower array 
\cite{Amenomori4}.
Shown in Fig.~\ref{fig:4} is the fraction of primary cosmic rays heavier 
than helium nuclei assuming the QGSJET model and the SIBYLL model 
which are  compared with those obtained recently by the KASCADE 
experiment \cite{Kampert}. 
Our results using 4 kinds of simulation models commonly
indicate the average mass of primary cosmic rays is 
going up around the knee, towards the direction of heavy dominance. 
On the other hand, the KASCADE results which measures both air shower 
size ($N_{\rm{e}}$) and muon size ($N_{\rm{\mu}}$) to deduce 
the energy spectrum of separate mass groups from the all-particle 
energy spectrum, strongly depend on the interaction models. 
The muon size contained in the air shower depends on the number of 
charged pions produced in the central and backward region 
(in the center of mass system) in the collisions of primary cosmic 
rays on air nuclei, which has a sizeable uncertainties experimentally 
as well as theoretically and is largely dependent on the interaction models.
From this point of view, the size of low-energy  muons $N_{\rm{\mu}}$ 
may not be a suitable parameter for separating the air showers into 
different primary mass groups.

\section{Summary}

 A hybrid experiment of emulsion chamber and  air-shower
array was successfully done at Yangbajing in Tibet to study the primary cosmic
 rays around the knee energy region.
Using the events observed simultaneously in the emulsion chamber and 
the air-shower array, and applying
a neural network analysis to this data set, we obtained the
energy spectrum of primary protons in the energy range from 4 $\times10^{14}$ eV 
to 10$^{16}$ eV.  The spectrum observed can be represented by the power-law fit 
and the power indexes are estimated to be $-3.01 \pm 0.11$ and $-3.05 \pm 0.12$ for the spectra obtained 
using the ANN trained by the QGSJET+HD and 
SIBYLL+HD events, respectively, which are steeper than that extrapolated 
from the direct observations of $-2.74 \pm 0.01$ in the energy range 
below $10^{14}$ eV. The absolute fluxes of protons was  derived 
within 30\% systematic errors depending on the hadronic interaction models
adopted in the Monte Carlo simulation. We also estimated  the primary helium
 spectrum 
at energies above 10$^{15}$ eV, which has almost same spectral slope with the
 proton spectrum. 

 We further obtained the result that the fraction of the nuclei heavier than
 helium in
 the primary cosmic rays around the knee region, which was estimated using 
the proton+helium spectrum  and the all-particle 
spectrum observed with the Tibet experiment, increases with increasing
primary energy. This strongly suggests that the main component responsible
for making the knee structure in the all-particle energy spectrum is
 the nuclei heavier than the helium component. 

This is the first measurement of the differential
energy spectra of primary protons and heliums by selecting them 
event by event. In the very near future, we will start a new high-statistics
 hybrid experiment in Tibet to clarify the main component of cosmic rays at the
knee. The new experiment is able to observe the air shower cores induced by
  heavy components around and beyond the knee, where direct
measurements are inaccessible because of their extremely low fluxes \cite{Xu}.

\section*{Acknowledgments}

 This work is supported in part by Grants-in-Aid for Scientific
Research on Priority Area (712) (MEXT) and also for Scientific Research (JSPS)
 in Japan,  and by the Committee of the Natural Science Foundation and by the
Chinese Academy of Sciences in China. The support of the JSPS (J.H., grant No. P03025)
is also acknowledged.


\end{document}